\documentclass[aps,epsfig,prl,twocolumn,superscriptaddress,showpacs,email,amssymb,nobibnotes]{revtex4}
\usepackage{graphicx,amsmath,amssymb,psfrag,epsfig,multirow,slashbox}
\begin{document}
\title{Stochastic effects in a seasonally forced epidemic model}
\author{G. Rozhnova}
\affiliation{Centro de F{\'\i}sica da Mat{\'e}ria Condensada and
Departamento de F{\'\i}sica, Faculdade de Ci{\^e}ncias da Universidade de
Lisboa, P-1649-003 Lisboa Codex, Portugal}
\affiliation{Departamento de Engenharia Civil, Instituto Superior de Engenharia de Lisboa, P-1959-007 Lisboa Codex, Portugal}
\author{A. Nunes}
\affiliation{Centro de F{\'\i}sica da Mat{\'e}ria Condensada and
Departamento de F{\'\i}sica, Faculdade de Ci{\^e}ncias da Universidade de
Lisboa, P-1649-003 Lisboa Codex, Portugal}
\begin{abstract}
The interplay of seasonality, the system's nonlinearities and intrinsic stochasticity
is studied for a seasonally forced susceptible-exposed-infective-recovered stochastic model. The model is explored in the parameter region that corresponds to childhood infectious diseases such as measles.
The power spectrum of the stochastic fluctuations around the attractors of the deterministic system that describes the model in the thermodynamic limit is computed analytically and validated by stochastic simulations for large system sizes. Size effects are studied through additional simulations. Other effects such as switching between coexisting attractors induced by stochasticity often mentioned in the literature as playing an important role in the dynamics of childhood infectious diseases are also investigated. The main conclusion is that
stochastic amplification, rather than these effects, is the key ingredient to understand the observed incidence patterns.
\end{abstract}
\maketitle

\section{I. Introduction}
Epidemic spread in human populations is a complex phenomenon whose comprehensive modeling 
has been a standing challenge for many years \cite{Keelingbook}. Several ingredients such as host population contact
structure, host heterogeneity, transmission mechanisms, interplay between immune response and pathogen evolution or 
demographic and environmental factors have been identified as playing an important role in 
the short and in the long term behavior of infection spread \cite{vespi, newman, girvan,dushoff,verdascaetal,Marder,Santos,brasileiros,boni,miller,nagy,chineses}. 
Scarcity of data, especially for long term behavior, and model parameters that are hard to measure accurately  add to the complexity of the problem, making it difficult to identify the key ingredients of a parsimonious model for a given disease or class of diseases \cite{KeelingGrenfellMeasles,NguyenRohani}.

Childhood infectious diseases have often been taken as a case study and model testing ground, because decades long of fairly well time resolved data records are available, on one hand, and because of their different phenomenology despite the similarities in contagion mechanisms and in infectious, latent and immunity waning typical times \cite{Anderson}. The common modeling approach for this class of diseases is a SEIR (susceptible-exposed-infective-recovered) compartmental model, with a periodic forcing that represents seasonal environmental or social factors that influence the transmission of the disease \cite{Anderson,schwartz1983,schwartz1984,schwartz1985,kuznetsov}. Deterministic models based on this approach, where the role of stochasticity is merely that of making the system switch between coexisting attractors, successfully reproduce the main features of the observed time series for measles incidence, but fail conspicuously to model the behavior of other childhood diseases  that exhibit non-seasonal sustained oscillations on several long term data records \cite{Earn01282000,BauchEarn}. This failure has been addressed in the literature by claiming that this different dynamics may be either the result of more realistic latent and infectious period distributions or the evidence of stochastic effects that would show up as noisy oscillations with the same frequency as the damped oscillations of the deterministic system \cite{BauchEarn,NguyenRohani}. The idea that stochastic effects may play a more 
fundamental role has also been discussed in the epidemiological literature \cite{RohaniKeelingGrenfell,KeelingGrenfellMeasles}, prompting several recent analytical studies, all of which deal with unforced systems \cite{DavidAlonso06222007,greenwood,Azaele}. 

Sustained oscillations typical of the incidence patterns of childhood infectious diseases are one of the features of the long term behavior 
of these unforced stochastic models. The power spectrum of the fluctuations around the endemic equilibrium computed analytically using van Kampen's system size expansion has well defined resonant like peaks \cite{DavidAlonso06222007}, which means that for moderate system sizes demographic stochasticity will generate sustained noisy oscillations, a phenomenon dubbed stochastic amplification when it was first studied in ecological and epidemiological models \cite{DavidAlonso06222007,McKane-Newman2005}.
Indeed, it has been shown that intrinsic noise enhanced by correlations in unforced epidemic models may give rise to oscillations that are comparable to those due to seasonal forcing in deterministic systems \cite{MSimoes05062008,PhysRevE.79.041922,PhysRevE.80.051915}. It has also been shown that the dominant frequency of these stochastic fluctuations may differ significantly from the characteristic frequency of the deterministic system, especially in the presence of correlations \cite{MSimoes05062008,PhysRevE.79.041922,andrea}. Adding seasonality to these unforced systems may also give rise to significant non trivial effects in their long term behavior. 
Therefore, the analytical results for the fluctuations power spectrum must be extended to the corresponding periodically forced models in order to assess the role of stochastic effects in childhood diseases epidemiology through the interplay between seasonality, the system's nonlinearities and intrinsic stochasticity. 

The method developed in Ref. \cite{McKane-Newman2005} was extended to fluctuations around cycles of forced or unforced systems in a series of recent papers \cite{unforcedmckane,forcedmckane}. Here, we apply this method to a seasonally forced SEIR model in a realistic parameter region for childhood infectious diseases where different attractors exist or coexist. Since very little is known about the amplitude of seasonal forcing, we leave this as a free parameter in our study and consider separately the low, intermediate and high forcing regimes. In all cases, we find an excellent agreement between the analytical power spectra and the results of stochastic simulations. We use the latter to assess the role of competing attractors in explaining the observed time series of childhood diseases epidemics, and we use the former to predict the number and position of the dominant non-seasonal peaks as a function of the epidemiological parameters.

\section{II. The seasonally forced deterministic SEIR epidemic model}
\label{seasdet}

In this section, we will review a seasonally forced deterministic SEIR model and its dynamics. 
Following most of the literature on childhood infectious diseases' modeling, we will take measles as a paradigmatic example throughout the paper, and the behavior of the system will be illustrated in the relevant parameter region \cite{schwartz1983,schwartz1985,KeelingGrenfellMeasles}. Extensions of the SIR/SEIR model have also been considered, especially in the mathematical literature, in connection with acute infectious diseases' modeling. These extensions may include, for instance, higher order nonlinearities in the infection term \cite{nonlinearity1,nonlinearity2,nonlinearity3}, immigration of infectives \cite{immigration,DavidAlonso06222007} or age structure \cite{Schenzle}. However, it is generally accepted that the seasonally forced SEIR model should capture the main features of the dynamics of childhood infections. 


Consider then a homogeneously mixed population of constant size consisting of four classes of individuals: susceptibles (healthy individuals capable of contracting the infection), exposed (infected but not yet infectious individuals), infectives (infectious individuals capable of transmitting the infection) and recovered (permanently immune individuals). The dynamics of the SEIR model is governed by the following processes: 

1) The susceptible individuals catch the infection from the infective individuals at a time dependent contact (or transmission) rate $\beta (t)$. To incorporate seasonality in transmission of the infection, the contact rate is assumed to be a periodic sinusoidal function with period 1 year 
\begin{equation}
\label{seasonalrate}
\beta(t)=\beta_0 (1+\beta_1 \cos 2\pi t),  
\end{equation} 
\noindent{}where $t$ is measured in years. In Eq. (\ref{seasonalrate}), $\beta_1 \in \left[0,1\right]$ is the amplitude of the seasonal variation in transmission and $\beta_0 > 0$ is the time-averaged contact rate. This form of the seasonal contact rate captures the first mode of periodic change in disease transmission due to the school terms and holidays \cite{comment}.   

2) The exposed individuals become infectious at the rate $\chi$. Hence, $1/\chi$ is the average latent period of the disease. 

3) The infective individuals recover at the rate $\gamma$ becoming permanently immune. The average infectious period of the disease is $1/\gamma$.

4) All individuals are subjected to the per capita death rate $\mu$ which is equal to the birth rate of the population. The average lifespan is given by $1/\mu$.

The seasonally forced deterministic SEIR model can now be written as follows 
\begin{eqnarray}
\label{seirforced1}
\dfrac{ds}{dt}&=&\mu\left(1-s(t)\right)-\beta_0 (1+\beta_1 \cos 2\pi t)s(t)i(t), \ \ \\
\dfrac{de}{dt}&=&\beta_0 (1+\beta_1 \cos 2\pi t)s(t)i(t)-(\chi+\mu)e(t),\\
\dfrac{di}{dt}&=&\chi e(t)-(\gamma+\mu)i(t),
\label{seirforced3}\end{eqnarray}       
\noindent{}where $s(t)$, $e(t)$ and $i(t)$ denote the fraction of susceptible, exposed and infective individuals, respectively. Note that the equation for the fraction of recovered individuals, $r(t)$, is redundant since $s(t)+e(t)+i(t)+r(t)=1$, where the size of the total population was normalized to unity.

The unforced deterministic SEIR model is recovered by setting $\beta_1=0$ \cite{Keelingbook,Anderson}: 
\begin{eqnarray}
\label{seirunforced1}
\dfrac{ds}{dt}&=&\mu\left(1-s(t)\right)-\beta_0 s(t)i(t),\\
\dfrac{de}{dt}&=&\beta_0 s(t)i(t)-(\chi+\mu)e(t),\\
\dfrac{di}{dt}&=&\chi e(t)-(\gamma+\mu)i(t).
\label{seirunforced3}\end{eqnarray}
\indent{}Both model (\ref{seirforced1})-(\ref{seirforced3}) and model (\ref{seirunforced1})-(\ref{seirunforced3}) have been extensively investigated in the literature \cite{Anderson,schwartz1983,schwartz1984,schwartz1985,kuznetsov}. Here, we will review well known facts about these models that are relevant for the present study. The dynamics of the unforced deterministic SEIR model (\ref{seirunforced1})-(\ref{seirunforced3}) depends on the basic reproductive ratio of the infection \cite{Keelingbook,Anderson}
\begin{equation}
\label{r0}
R_0=\frac{\beta_0\chi}{ (\chi+\mu)(\gamma+\mu)}, 
\end{equation}
\noindent{}defined as the average number of secondary cases caused by an infectious individual in a totally susceptible population in one infectious period. The stability analysis shows that Eqs. (\ref{seirunforced1})-(\ref{seirunforced3}) have an asymptotically stable endemic equilibrium 
\begin{equation}
\label{endemicseir}
\left(s^*,e^*,i^*\right)=\left(\dfrac{1}{R_0},\dfrac{\mu(\gamma+\mu)(R_0-1)}{\beta_0\chi},\dfrac{\mu(R_0-1)}{\beta_0}\right), 
\end{equation}
\noindent{}provided $R_0>1$. The trivial steady state  
\begin{equation}
\label{trivialseir}
\left(s^*,e^*,i^*\right)=(1,0,0) 
\end{equation}
\noindent{}is stable for $R_0<1$ and unstable for $R_0>1$. 

\begin{table}[t]
\centering
{\begin{tabular}{|c|c|c|}
\hline \multicolumn{3}{|c|}{Fixed parameters} \\ 
\hline \ \ \ \ \ Rate of disease onset \ \ \ \ \ & \ \ \ $\chi$ \ \ \  & \ \ \ $35.84$ $\text{year}^{-1}$ \ \ \ \\ 
\hline Average latent period & \ \ \ $1/\chi$ \ \ \ & \ \ \ $10.18$ days \ \ \ \\
\hline Rate of recovery  & \ \ \ $\gamma$  \ \ \ & \ \ \ $100$ $\text{year}^{-1}$ \ \ \ \\
\hline Average infectious period & \ \ \ $1/\gamma$ \ \ \ & \ \ \ $3.65$ days \ \ \ \\
\hline Birth/death rate & \ \ \ $\mu$ \ \ \ & \ \ \ $0.02$ $\text{year}^{-1}$ \ \ \ \\  
\hline Average lifespan & \ \ \ $1/\mu$ \ \ \ & \ \ \ $50$ years \ \ \ \\     
\hline Average contact rate  & \ \ \ $\beta_0$  \ \ \ & \ \ \ $1575$ $\text{year}^{-1}$ \ \ \ \\
\hline Basic reproductive ratio & \ \ \ $R_0$ \ \ \ & \ \ \ $15.74$ \ \ \ \\
\hline \multicolumn{3}{|c|}{Variable parameter} \\
\hline   &  &  $0.02$ ($T=1$) \\
  &  &  $0.05$ ($T=1$) \\
\ Amplitude of seasonal forcing \ &  \ \ \ $\beta_1$  \ \ \ &  $0.10$ ($T=1,3$) \\
   &  &  $0.12$ ($T=2,3$) \\
    &  &  $0.2$ ($T=2$) \\
\hline
\end{tabular}
\caption{Parameter values for measles that will be used in this paper. According to Eq. (\ref{r0}) this set corresponds to $R_0= 15.74$. The amplitude of seasonal forcing will be varied so that the solutions of Eqs. (\ref{seirforced1})-(\ref{seirforced3}) exhibit stable limit cycles of period $T$ indicated in the parenthesis next to the $\beta_1$ value.}
\label{table4}}
\end{table}

If the contact rate varies seasonally according to Eq. (\ref{seasonalrate}) then a wide range of dynamical behavior of the seasonally forced deterministic SEIR model (\ref{seirforced1})-(\ref{seirforced3}) is possible, depending on $R_0$ and on the amplitude of seasonal forcing $\beta_1$. In what follows, we will restrict our analysis to the case of measles and use the parameter values given in Refs. \cite{schwartz1983,schwartz1985}. The birth/death rate will be fixed at $0.02$ $\text{year}^{-1}$ which corresponds to the average lifespan of $50$ years. The average latent and infectious periods will be equal to $10.18$ days and $3.65$ days, respectively, yielding $13.83$ days for the average time interval between infection and recovery. These values correspond to $35.84$ $\text{year}^{-1}$ for the disease onset rate and $100$ $\text{year}^{-1}$ for the  recovery rate. Finally, the average contact rate is adjusted to be $1575$ $\text{year}^{-1}$ so that the basic reproductive ratio is $15.74$. The remaining parameter, the forcing amplitude, will be given different values in the interval $\left[0.02,0.2\right]$ (estimates for $\beta_1$ can be found in Ref. \cite{KeelingGrenfellMeasles}). The summary of the parameter values is shown in Table \ref{table4}.  

The bifurcation analysis of the seasonally forced deterministic SEIR model (\ref{seirforced1})-(\ref{seirforced3}) for the fixed set of parameter values considered above and $\beta_1$ as a single free parameter can be found in Refs. \cite{schwartz1983,schwartz1985}. In Ref. \cite{schwartz1984} an analysis of the same model was performed for variable $\beta_1$ and $R_0= 18$ (corresponding to $\beta_0=1800$ $\text{year}^{-1}$). For the interested reader, we recommend Ref. \cite{kuznetsov} where more general bifurcation diagrams of Eqs. (\ref{seirforced1})-(\ref{seirforced3}) were computed with two free parameters $R_0$ and $\beta_1$ and the remaining parameters held constant as in Table \ref{table4}. 

For the parameter values of Table \ref{table4} and variable $\beta_1$, a brief summary of the bifurcation diagram is as follows \cite{schwartz1983,schwartz1985}. If $\beta_1$ is positive but small, a stable limit cycle of period 1 bifurcates from the endemic equilibrium point (\ref{endemicseir}). As $\beta_1$ is increased monotonically, it is found that at a value of $\beta_1'\approx 0.11479$ a stable branch of period 2 bifurcates from the period 1 branch, and the period 1 branch becomes unstable for $\beta_1>\beta_1'$. Additionally, in some range of $\beta_1$ these period 1 and period 2 limit cycles coexist with a pair of limit cycles of period 3 (one stable and one unstable) appearing from a saddle-node bifurcation. Finally, in a very narrow window of $\beta_1$ the period 2 and period 3 branches exhibit a cascade of period-doubling bifurcations as $\beta_1$ increases, leading to chaotic epidemics. The full bifurcation diagram contains yet another stable attractor of period 4 that coexists with stable period 2 and period 3 limit cycles, however this attractor has a very small basin of attraction and it is hard to spot both in numerical integrations of Eqs. (\ref{seirforced1})-(\ref{seirforced3}) and in stochastic simulations.     

Thus, in the realistic parameter region there are three main stable attractors, namely stable limit cycles of period 1, 2 and 3. We will consider typical values of $\beta_1$ for which i) only a limit cycle of period 1 exists ($\beta_1 = 0.02, 0.05$); ii) limit cycles of period 1 and 3 coexist ($\beta_1 =0.1$); iii) limit cycles of period 2 and 3 coexist ($\beta_1 =0.12$); iv) a limit cycle of period 2 exists ($\beta_1=0.2$). From now on the reader can refer to Table \ref{table4} where the periods $T$ of the (co)existing limit cycles are indicated for each value of $\beta_1$. Note that the regimes ii) and iii) lie in the vicinity of the bifurcation point where a stable limit cycle of period 2 is born ($\beta_1'\approx 0.11479$). 

\begin{figure}
\centering
{\includegraphics[trim=2.8cm 0.4cm 4.95cm 0.cm, clip=true, width=\columnwidth]{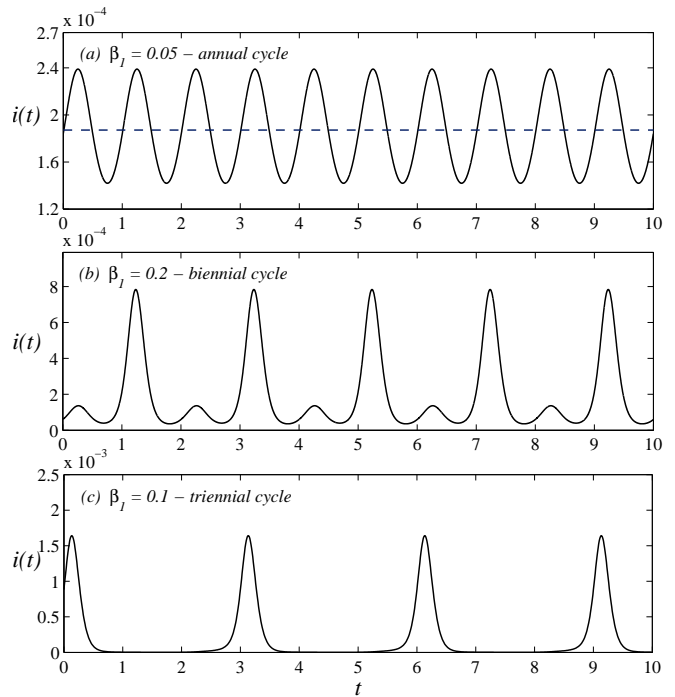}
\caption{The fraction of infective individuals as a function of time for periodic solutions of Eqs. (\ref{seirforced1})-(\ref{seirforced3}).} 
\label{chap6fig1}} 
\end{figure}     

The fraction of infectives for stable periodic solutions of the seasonally forced SEIR model (\ref{seirforced1})-(\ref{seirforced3}) is shown in Fig. \ref{chap6fig1}. The solid curves were computed by numerical integration of Eqs. (\ref{seirforced1})-(\ref{seirforced3}) for the parameters values given in Table \ref{table4} and three values of the forcing amplitude. In all cases, initial conditions were chosen on a cycle so as to avoid transients. The annual epidemic cycle [$\beta_1=0.05$, Fig. \ref{chap6fig1} (a)] corresponds to a small amplitude stable limit cycle of period 1 bifurcating from the endemic equilibrium (\ref{endemicseir}) of the unforced SEIR model (\ref{seirunforced1})-(\ref{seirunforced3}). This cycle is present in the system if $0<\beta_1 < 0.11479$. The equilibrium value of infectives, $i^*$, in the unforced model (\ref{seirunforced1})-(\ref{seirunforced3}) is also shown (the dashed line). The biennial epidemic cycle representing the alternating years of high and low incidence can be found for $0.11479 <\beta_1 \leq 0.2$ (here we do not consider values of $\beta_1>0.2$ where a stable limit cycle 
of period 2 is still present because such levels of forcing are considered unrealistically high). A plot of typical biennial epidemic is shown for $\beta_1=0.2$ in Fig. \ref{chap6fig1} (b). Such a behavior of measles is in accordance with the data from the New York City and from England and Wales recorded between 1950 and 1968 (before vaccination) \cite{Schenzle,Anderson,KeelingGrenfellMeasles}. As for other stable attractors coexisting with stable limit cycles of period 1 and 2 these are, mainly, limit cycles of period 3 as mentioned in this section before. A typical limit cycle of period 3 is characterized by higher amplitude outbreaks and by very low minima of infective incidence, see Fig. \ref{chap6fig1} (c) where the behavior of the solution for the fraction of infectives is shown for $\beta_1=0.1$.

\section{III. The seasonally forced stochastic epidemic model}
In the seasonally forced stochastic SEIR model, a discrete population of the constant size $N$ is divided into the classes of susceptibles ($S$), exposed ($E$), infectives ($I)$ and recovered ($R$). At any instant of time, the total number of individuals equals $N$ thus only three of the classes are independent. Denoting numbers of susceptible, exposed and infective individuals by $m_1, m_2$ and $m_3$, respectively, and considering the processes postulated in Section II, we can obtain the transition rates for the stochastic model as follows. 

1) The infection process takes place between a susceptible and an infective individual at the time dependent contact rate $\beta(t)$ and results in the transition of the susceptible to the exposed class, $S+I\stackrel{\beta(t)}{\longrightarrow} E+I$. The corresponding transition rate is
\begin{equation}
\label{StoE}
\mathcal{T}^{m_1-1,m_2+1,m_3}_{m_1,m_2,m_3}=\beta_0 (1+\beta_1\cos 2\pi t)\frac{m_1}{N}m_3,  
\end{equation}
\noindent{}where the superscript and the subscript of $\mathcal{T}$ denote the final and the initial states of the system.

2) The transition of an individual from the exposed class to the infective class, $E\stackrel{\chi}{\longrightarrow} I$, occurs at the rate $\chi$. The transition rate of this process equals
\begin{equation}
\label{EtoI}
\mathcal{T}^{m_1,m_2-1,m_3+1}_{m_1,m_2,m_3}=\chi m_2. 
\end{equation}
\indent{}3) The transition rate of the recovery of an infective individual occurring at the rate $\gamma$, $I\stackrel{\gamma}{\longrightarrow} R$, is given by 
\begin{equation}
\label{ItoR}
\mathcal{T}^{m_1,m_2,m_3-1}_{m_1,m_2,m_3}=\gamma m_3 . 
\end{equation}
\noindent{}The recovered individuals are permanently immune.

4) Finally, there are four transition rates associated with the linked birth, $R\stackrel{\mu}{\longrightarrow}S$, and death processes, $S\stackrel{\mu}{\longrightarrow} R$, $E\stackrel{\mu}{\longrightarrow}R$ and $I\stackrel{\mu}{\longrightarrow}R$, occurring at the rate $\mu$:
\begin{eqnarray}
\mathcal{T}^{m_1+1,m_2,m_3}_{m_1,m_2,m_3}&=&\mu N , \\
\mathcal{T}^{m_1-1,m_2,m_3}_{m_1,m_2,m_3}&=&\mu m_1, \\
\mathcal{T}^{m_1,m_2-1,m_3}_{m_1,m_2,m_3}&=&\mu m_2, \\
\label{mus} 
\mathcal{T'}^{m_1,m_2,m_3-1}_{m_1,m_2,m_3}&=&\mu m_3. 
\end{eqnarray}
\noindent{}Note that the initial and final states for the recovery of an infective individual and the death of an infective individual are equal, see Eqs. (\ref{ItoR}) and (\ref{mus}) but the corresponding transition rates are different. To distinguish between the latter we use a prime in Eq. (\ref{mus}). 

The dynamics of the stochastic system determined by Eqs. (\ref{StoE})-(\ref{mus}) is completely described by the master equation \cite{vanKampen,Risken}. Given the initial and boundary conditions, this equation expresses the evolution of the probability of having a system in state with $m_1$ susceptibles, $m_2$ exposed and $m_3$ infectives, $\mathcal{P}(m,t)$, at any positive time. Denoting $m$ as the shorthand for three numbers $m_1,m_2,m_3$ this equation is written in the following form:
\begin{equation}
\label{generalmasterequation}\dfrac{d\mathcal{P}(m,t)}{dt}=\sum\limits_{m'\neq m} \Big[\mathcal{T}^{m}_{m'}\mathcal{P}(m',t)-\mathcal{T}^{m'}_{m}\mathcal{P}(m,t) \Big] .
\end{equation}
\noindent{}The positive term on the r.h.s. of Eq. (\ref{generalmasterequation}) is the increase of $\mathcal{P}(m,t)$ due to the transitions from states $m'$ to state $m$, while the negative term is the decrease due to the transitions from state $m$ to states $m'$. The term with $m$ equal to $m'$ is excluded because its contribution is zero. 

Substituting Eqs. (\ref{StoE})-(\ref{mus}) into Eq. (\ref{generalmasterequation}) one obtains the master equation for the seasonally forced stochastic SEIR model (\ref{masterseasonal}) (see the Appendix). This equation is in fact a system of $m_1\cdot m_2\cdot m_3$ coupled differential-difference equations similar to those used in studies of simple birth-death stochastic processes \cite{Risken,Goel}. In that case, an exact solution can be found in terms of generating functions or by other means but Eq. (\ref{masterseasonal}) with three variables $m_1$, $m_2$ and $m_3$ is too complicated to be solved exactly. To get insight into the dynamics of the seasonally forced stochastic SEIR model for large system sizes, we will apply a method due to van Kampen \cite{vanKampen}. Similarly to other studies \cite{unforcedmckane,forcedmckane}, we will expand Eq. (\ref{masterseasonal}) in powers of the inverse of the system size $N$. First, let us rewrite Eq. (\ref{masterseasonal}) in terms of the step operators $\epsilon^{\pm 1}_j$, where $j=1,2,3$, defined by their action on a smooth function $f(m_1,m_2,m_3,t)$ \cite{vanKampen}:      
\begin{eqnarray}
\label{epsilon1seas}
\epsilon^{\pm 1}_1 f(m_1,m_2,m_3,t)&=&f(m_1\pm 1,m_2,m_3,t), \ \  \\
\epsilon^{\pm 1}_2 f(m_1,m_2,m_3,t)&=&f(m_1,m_2\pm 1,m_3,t),\\
\label{epsilon3seas}
\epsilon^{\pm 1}_3 f(m_1,m_2,m_3,t)&=&f(m_1,m_2,m_3\pm 1,t). 
\end{eqnarray} 
\noindent{}Using Eqs. (\ref{epsilon1seas})-(\ref{epsilon3seas}) the master Eq. (\ref{masterseasonal}) transforms into Eq. (\ref{seasmastlast}) given in the Appendix. 

For the seasonally forced stochastic SEIR model the expansion is made around a deterministic periodic solution of Eqs. (\ref{seirforced1})-(\ref{seirforced3}) which is a stable limit cycle of period $T$ (see Ref. \cite{forcedmckane}). Thus, we make a transformation from the discrete variables $m_j(t)$ to the continuous variables $x_j(t)$, where $j=1,2,3$, according to the equations
\begin{eqnarray}
\label{tr1seas}
m_1(t)&=&N \bar{s}(t) + \sqrt{N} x_1(t),\\
m_2(t)&=&N \bar{e}(t) + \sqrt{N} x_2(t),\\
\label{tr3seas}
m_3(t)&=&N \bar{i}(t) + \sqrt{N} x_3(t),
\end{eqnarray}
\noindent{}where $\bar{s}(t), \bar{e}(t)$ and $\bar{i}(t)$ denote the deterministic trajectory of Eqs. (\ref{seirforced1})-(\ref{seirforced3}) and $x_1(t), x_2(t)$ and $x_3(t)$ are the corresponding stochastic fluctuations about it. After the substitution of Eqs. (\ref{StoE})-(\ref{mus}) and Eqs. (\ref{tr1seas})-(\ref{tr3seas}) into Eq. (\ref{seasmastlast}), the terms of different orders in $N$ can be identified in Eq. (\ref{seasmastlast}). The leading order terms yield Eqs. (\ref{seirforced1})-(\ref{seirforced3}) with  the substitutions $s(t)=\bar{s}(t),e(t)=\bar{e}(t),i(t)=\bar{i}(t)$:
\begin{eqnarray}
\dfrac{d\bar{s}}{dt}&=&\mu\left(1-\bar{s}(t)\right)-\beta_0 (1+\beta_1 \cos 2\pi t)\bar{s}(t)\bar{i}(t), \ \ \\
\dfrac{d\bar{e}}{dt}&=&\beta_0 (1+\beta_1 \cos 2\pi t)\bar{s}(t)\bar{i}(t)-(\chi+\mu)\bar{e}(t),\\
\dfrac{d\bar{i}}{dt}&=&\chi \bar{e}(t)-(\gamma+\mu)\bar{i}(t).
\end{eqnarray}   
\noindent{}The next-to-leading order terms yield a multivariate linear Fokker-Plank equation for the probability distribution $\Pi(x,t)$ \cite{vanKampen,Risken}
\begin{equation}
\label{fokkerplanckseas}
\dfrac{\partial\Pi}{\partial t}=-\sum\limits_{j,k}{A_{jk}(t)\dfrac{\partial (x_k\Pi)}{\partial x_j}}+\frac{1}{2}\sum\limits_{j,k}{B_{jk}(t)\dfrac{{\partial}^2\Pi}{\partial x_j \partial x_k}} , 
\end{equation}
\noindent{}where $j,k=1,2,3$. In Eq. (\ref{fokkerplanckseas}), ${\bf A}(t)$ is the Jacobian of Eqs. (\ref{seirforced1})-(\ref{seirforced3})
\begin{equation}
\label{Aseir}
{\bf A}(t)= \left( \begin{array}{ccc} -\mu-\beta(t)\bar{i}(t) & 0 & -\beta(t)\bar{s}(t) \\
 \beta(t)\bar{i}(t) & -(\chi+\mu) & \beta(t)\bar{s}(t) \\
0 & \chi & -(\gamma+\mu) \end{array} \right) 
\end{equation}
\noindent{}and ${\bf B}(t)$ is the symmetric cross correlation matrix
\begin{equation}
\label{Bseir}
{\bf B}(t)= \left( \begin{array}{ccc} \ \ f_{11}(t) & -f_{12}(t) & 0 \\
-f_{12}(t) & \ \ f_{22}(t)& -f_{23}(t) \\
0 & -f_{23}(t) & \ \ f_{33}(t) \end{array} \right) ,  
\end{equation} 
\noindent{}where 
\begin{eqnarray}
f_{11}(t)&=&\mu(1+\bar{s}(t))+f_{12}(t), \\
f_{22}(t)&=&\mu \bar{e}(t)+f_{12}(t)+f_{23}(t), \\
f_{33}(t)&=&(\gamma+\mu)\bar{i}(t)+f_{23}(t)
\end{eqnarray} 
\noindent{}and 
\begin{eqnarray}
f_{12}(t)&=&\beta(t)\bar{s}(t)\bar{i}(t), \\
f_{23}(t)&=&\chi \bar{e}(t) .
\end{eqnarray}
\noindent{}Both matrices ${\bf A}(t)$ and ${\bf B}(t)$ are evaluated on the limit cycle solutions of Eqs. (\ref{seirforced1})-(\ref{seirforced3}) and thus they are periodic functions of $t$ with the same period $T$ of the limit cycle,
\begin{equation}
\label{per}
{\bf A}(t)={\bf A}(t+T), \ \ {\bf B}(t)={\bf B}(t+T).
\end{equation}   
\indent{}In previous studies of unforced epidemic models \cite{PhysRevE.79.041922,PhysRevE.80.051915,MSimoes05062008,andrea}, the power spectrum of stochastic fluctuations $x_j(t)$ was computed from the multivariate Langevin equation associated with Eq. (\ref{fokkerplanckseas}) \cite{vanKampen,Risken}
\begin{equation}
\label{langevinseir}
\dfrac{dx_j(t)}{dt}=\sum\limits_k{A_{jk} (t) x_k(t)}+L_j(t), \ \ j,k=1,2,3,
\end{equation}
\noindent{}where $L_j(t)$ are Gaussian noise terms with zero mean 
\begin{equation}
          \left\langle L_j(t)\right\rangle =0 
\end{equation}
\noindent{}and with the correlator
\begin{equation}
\left\langle L_j(t)L_k(t')\right\rangle =B_{jk}(t)\delta(t-t') .
\end{equation}
\noindent{}The analytical calculation of the power spectrum through the Fourier transform of Eq. (\ref{langevinseir}) done in those studies depends on the fact that the matrices ${\bf A}$ and ${\bf B}$ are constant in the unforced case. For the seasonally forced stochastic SEIR model, the matrices ${\bf A}(t)$ and ${\bf B}(t)$ are time dependent and this method does not work anymore. However, one can use the periodicity of ${\bf A}(t)$ and ${\bf B}(t)$ and Floquet's theory to find a solution of Eq. (\ref{langevinseir}) and compute its power spectrum, as briefly outlined below. 
This method was developed to study the effects of external noise in nonlinear oscillations close to bifurcations \cite{wiesenfeld, wiesenfeld2}, and it has been applied to intrinsic stochasticity for several systems similar to the model considered here \cite{forcedmckane}. 

The general solution of the inhomogeneous system of first-order linear differential equations (\ref{langevinseir}) with matrix function ${\bf A}(t)$ and vector function ${\bf L}(t)$ can be written as a sum of the general solution of the homogeneous system
\begin{equation}
\label{homseir}
\dfrac{dx_j(t)}{dt}=\sum\limits_k{A_{jk} (t) x_k(t)}, \ \ j,k=1,2,3,
\end{equation}
\noindent{}and a particular solution of the inhomogeneous system \cite{Grimshaw}. Furthermore, the general solution of the homogeneous system (\ref{homseir}) with ${\bf A}(t)$ periodic with period $T$ obeys Floquet's theorem \cite{Grimshaw,Meirovitch}.  This theorem states that if $\textbf{X}(t)$ is a fundamental matrix of Eq. (\ref{homseir}), then 
there exists a periodic nonsingular matrix $\textbf{Q}(t)$ with period $T$ and a constant matrix $\textbf{R}$ such that 
\begin{equation}
\textbf{X}(t)=\textbf{Q}(t)e^{t\textbf{R}}, \ \ \textbf{Q}(t)=\textbf{Q}(t+T) .
\end{equation} 
\noindent{}The matrix $\textbf{D}=e^{T\textbf{R}}$ is sometimes referred to as the monodromy matrix of the fundamental matrix $\textbf{X}(t)$. Although $\textbf{D}$ is not unique, its eigenvalues, $\lambda_1 , \lambda_2 , \lambda_3$, called the characteristic (Floquet) multipliers associated with the periodic matrix $\textbf{A}(t)$, are unique. The eigenvalues of matrix $\textbf{R}$, $\rho_1 , \rho_2 , \rho_3$, are called the characteristic (Floquet) exponents associated with the periodic matrix $\textbf{A}(t)$. The latter are related to the characteristic multipliers by 
\begin{equation}
\rho_j=\frac{1}{T}\log| \lambda_j|, \ \ j=1,2,3,
\end{equation} 
\noindent{}where the principal value of the logarithm is taken. 

Using further Floquet's theory an analytical expression can be obtained for the autocorrelation function of the stochastic fluctuations 
\begin{equation}
\label{corr}
C_{jj}(\tau)=\frac{1}{T}\int\limits_0^T \langle x_j (t) x_j (t+\tau) \rangle \text{d}t , \ \ j=1,2,3,
\end{equation} 
\noindent{}in terms of the matrices $\textbf{Q}(t), \textbf{R}$ and $\textbf{B}(t)$ \cite{forcedmckane,wiesenfeld}. Note that Eq. (\ref{corr}) does not depend on the initial condition if the initialization time is set to the infinite past. The power spectrum (power spectral density) $\textbf{P}_j(\omega)$ of the stochastic fluctuations can then be computed from the autocorrelation function $C_{jj}(\tau)$ via the Fourier transform (the Wiener\text{-}Khinchin theorem) 
\begin{equation}
\textbf{P}_j(\omega)=\int\limits_{-\infty}^{+\infty} C_{jj}(\tau) e^{-\text{i}\omega \tau} \text{d}\tau .
\end{equation}
\indent{}In the following section, we will compare the power spectra calculated analytically 
through the method described above
with those measured from simulations of the stochastic process defined by Eqs. (\ref{StoE})-(\ref{mus}). The spectra of the fluctuations of infectives are of particular interest to us because they can be compared with the spectra computed from real incidence data \cite{BauchEarn,Bausch}.

\section{IV. Results}

The stochastic process is simulated with the use of a modified Gillespie algorithm \cite{gillespie} to account for the explicit time dependence in the transition rates \cite{DavidAnderson,Lu}. Unless explicitly stated otherwise, all simulations start from a random initial condition and time series of 500 years or 100 years are recorded (50 or 10 years of which are considered as transient). By definition the model does not include injection of infectives, thus if an extinction occurs before this time a simulation is discarded. The largest system size we tested is $N=10^8$ and the smallest one is $N=5\times 10^5$. For some amplitudes of seasonal forcing the number of extinctions gets huge and we were not able to run such long simulations for system sizes smaller than $N=10^6$. However, shorter time series could have been obtained (for comparison with the prevaccination era these should be about 20 years long \cite{KeelingGrenfellMeasles}). 

\begin{table}
\centering
{\begin{tabular}{|c|c|c|c|}
\hline \ \ \ $\beta_1$ \ \ \ & \ \ $T$ \ \ & \ \ Floquet multipliers \ \ & \ \ Floquet exponents \ \ \\
\hline \multirow{2}{*}{$0.02$} & \multirow{2}{*}{1} & $-0.8262\pm \text{i} \ 0.3007 $ & $-0.1287\pm \text{i} \ 2.7926$ \\
  & & $9.1867\times 10^{-60}$ & $-135.9373$ \\ 
\hline \multirow{2}{*}{$0.05$} & \multirow{2}{*}{1} & $-0.8369\pm \text{i} \ 0.2696$ & $-0.1287\pm \text{i} \ 2.8299$ \\
  & & $9.1868\times 10^{-60}$ & $-135.9373$ \\
\hline \multirow{4}{*}{$0.10$} & \multirow{2}{*}{1} & $-0.8724\pm \text{i} \ 0.1091$ & $-0.1287\pm \text{i} \ 3.0172$ \\
  & & $9.1866\times 10^{-60}$ & $-135.9374$ \\
\cline{2-4}  & \multirow{2}{*}{3} & $-0.2715\pm \text{i} \ 0.6119$ & $-0.1338 \pm \text{i} \ 0.6628$ \\
 &  & $7.7134 \times 10^{-178}$ & $-135.9391$ \\ 
\hline \multirow{4}{*}{$0.12$} & \multirow{2}{*}{2} & $0.7618 \pm \text{i} \ 0.1310$ & $-0.1287\pm \text{i} \ 0.0852$ \\
  & & $8.4390\times 10^{-119}$ & $-135.9374$ \\
\cline{2-4}  & \multirow{2}{*}{3} & $-0.6562\pm \text{i} \ 0.1305$ & $-0.1340 \pm \text{i} \ 0.9878$ \\
 &  & $7.7122 \times 10^{-178}$ & $-135.9391$ \\ 
\hline \multirow{2}{*}{$0.2$} & \multirow{2}{*}{2} & $0.1184\pm \text{i} \ 0.7630$ & $-0.1293 \pm \text{i} \ 0.7085$ \\
   & & $8.4357\times 10^{-119}$ & $-135.9376$ \\
\hline
\end{tabular}
\caption{Floquet multipliers, $\lambda_1, \lambda_2, \lambda_3$, and Floquet exponents, $\rho_1, \rho_2, \rho_3$, for limits cycles of different periods and several values of the forcing amplitude. All other parameters are kept fixed as in Table \ref{table4}.}
\label{table5}}
\end{table}

There is one difficulty in the computation of the analytical autocorrelation functions and analytical power spectra. Although an explicit expression for the autocorrelation function can be found, the final results have to be obtained numerically because the stable limit cycle is not known in closed form. In the endemic equilibrium, the eigenvalues of the Jacobian of the unforced SEIR deterministic model (\ref{seirunforced1})-(\ref{seirunforced3}) differ by two orders of magnitude for the parameter values typical of childhood infectious diseases.
This is reflected in the seasonally forced deterministic SEIR model (\ref{seirforced1})-(\ref{seirforced3}) \cite{schwartz1983} in the same parameter region. For the stable limit cycles given in Table \ref{table4}, two of the characteristic multipliers are always complex conjugate and have real part of order $1$, while the third one, $\lambda_3$, is of order $10^{-59}$, $10^{-118}$ and $10^{-177}$ for limit cycles of periods 1, 2 and 3, respectively. In this way, the monodromy matrix $\textbf{D}$ becomes singular ($\det \textbf{D} = \lambda_1 \lambda_3 \lambda_3$) in double precision numerical calculations and the periodic matrix $\textbf{Q}(t)$ cannot be computed. 
One way to circumvent this difficulty is to reduce the computations to the two dimensional central manifold associated with the complex eigenvalues, disregarding the dynamics along the strongly stable manifold associated with the small real eigenvalue $\lambda_3$ where fluctuations will be strongly damped (a similar approach was used heuristically in Ref. \cite{BauchEarn}). Here we adopt a direct approach by implementing a Runge-Kutta 7-8 method for the numerical integration of the differential equations on the limit cycle and by using arbitrary precision libraries \textit{NTL} and \textit{GMP} \cite{ntlgmp}. Typically the working precision set up in the integrations was 5 digits higher than the smallest characteristic multiplier (for example, 65 digits for limit cycles of period 1) and the numerical trajectory was correct up to 50 digits. Thus we could perform all the computations in the original variables of the full three dimensional system, which allows an immediate comparison with the power spectra of the simulations. The summary of the Floquet multipliers and Floquet exponents computed in this way is given in Table \ref{table5}. 

Note that until now we have not raised the question of the stability of the deterministic limit cycles because we do know from previous studies that the cycles are stable \cite{schwartz1983,schwartz1985,kuznetsov}. However, we automatically check this by computing the absolute values of $\lambda_j$ which must be less than unity for a limit cycle to be asymptotically stable (or, alternatively, the real parts of $\rho_j$ must be negative) \cite{Grimshaw,Meirovitch}. The homogeneous equation (\ref{homseir}) is, in fact, a variational equation for small perturbations $x_j(t)$ around the periodic solution of Eqs. (\ref{seirforced1})-(\ref{seirforced3}) \cite{Grimshaw,Meirovitch}. The presence of two complex conjugate Floquet multipliers implies that deterministic perturbations decay to the cycle in a damped oscillatory way. This situation is similar to that of unforced systems with a stable focus in which resonant amplification of stochastic fluctuations was observed \cite{McKane-Newman2005,PhysRevE.79.041922}. It will become clear in what follows that in the stochastic system with seasonal forcing fluctuations around the limit cycles are significantly amplified too, and that the analytical and simulated power spectra cannot be explained by the deterministic theory alone. 

\begin{figure}
\centering
{\includegraphics[trim=2.1cm 3.1cm 5.5cm 0.1cm, clip=true, width=\columnwidth]{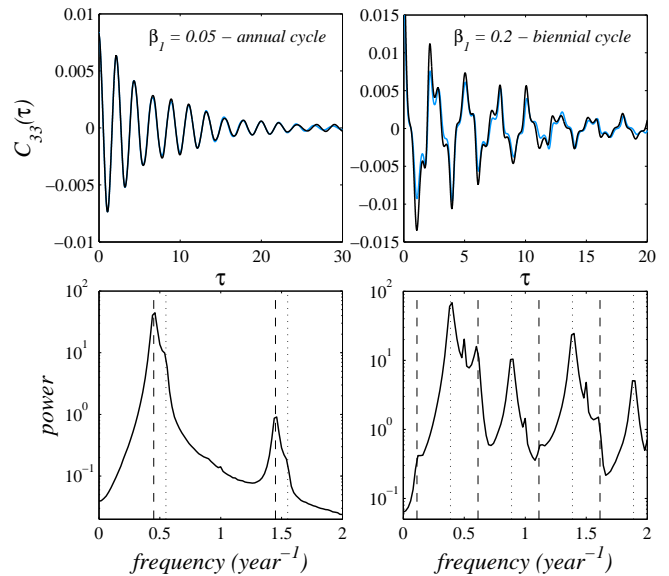}
\vspace{0.1cm}
\caption{(Color online) The upper panels show the autocorrelation functions of the stochastic fluctuations of infectives. The theoretical curves are plotted in gray (blue), and the curved computed from the simulations are plotted in black (for the annual cycle the analytical and simulated autocorrelation functions are strictly coincident). The lower panels show the corresponding power spectra. The vertical helper lines mark the frequencies predicted by Eq. (\ref{fr}). The dashes (dotted) lines are calculated by taking the plus (minus) in the equation. The system size used for simulation is $10^8$.} \label{chap6fig2}} 
\end{figure} 

To validate the theory developed in this study, we first compare analytical and simulated autocorrelation functions for the stochastic fluctuations of infectives. Typical values of the forcing amplitude are chosen in the annual and biennial regime (see left and right upper plots of Fig. \ref{chap6fig2}). Far from bifurcation points and for large system sizes, we find an excellent agreement in both cases (a small divergence can occur because of the sparse discretization of the orbit which we are lead to do to avoid very heavy computations). The lower plots of Fig. \ref{chap6fig2} show the corresponding power spectra. More exactly, the $x$-axes stands for temporal frequency and the $y$-axes is the power of the discrete Fourier transform of the autocorrelation function time series. In this and in the following plots, the power spectra are normalized so that the total power is the summed squared amplitude of the time series \cite{numericalrecipes}. One observes that the power spectra exhibit a number of peaks occurring regularly. As it has been noticed before by several authors \cite{wiesenfeld,forcedmckane}, the peaks are located at frequencies 
\begin{equation}
\nu_n=\frac{n}{T}\pm \frac{|\text{Im} \ {\rho_{1,2}}|}{2\pi},
\label{fr}
\end{equation}  
\noindent{}where $n=0,\pm 1, \pm 2, \ldots$, $|\text{Im} \ {\rho_{1,2}}|$ denotes the absolute value of the imaginary part of complex conjugate Floquet exponents (see Table \ref{table5}), and $T$ is period of a limit cycle. Note that for the annual limit cycle the main stochastic peak is situated at $|\text{Im} \ {\rho_{1,2}}|/(2\pi)$ while for the biennial cycle this peak is much smaller than the peak at $1/2-|\text{Im} \ {\rho_{1,2}}|/(2\pi)$ which is dominant. The plots are done in lin-log scale for a better observation of the structure of the power spectra.

\begin{figure}
\centering
{\includegraphics[trim=3.7cm 0.05cm 4.05cm 0.15cm, clip=true, width=\columnwidth]{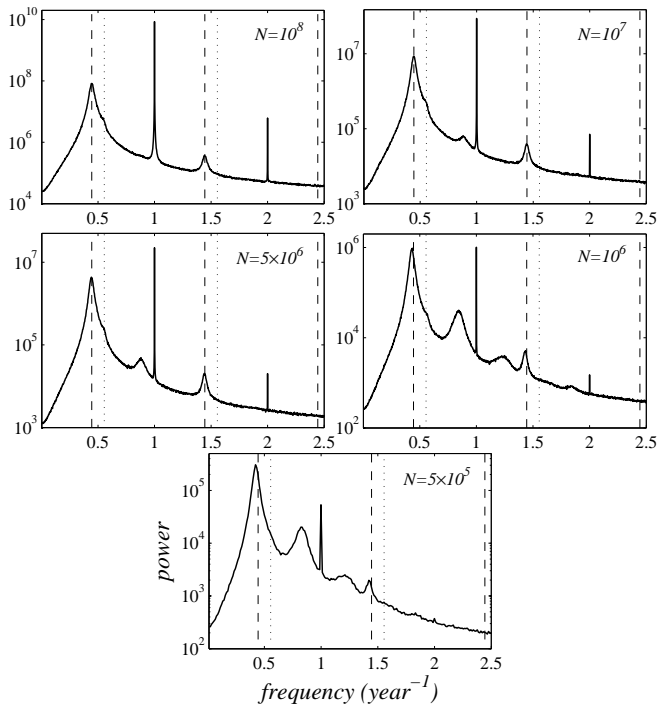}
\vspace{0.1cm}
\caption{Power spectra of the number of infectives calculated from simulations for several system sizes $N$. The simulations of 500 years were run for $N=10^8, 10^7, 5\times10^6, 10^6$ and of 100 years for $N=5\times10^5$. Averages over $10^3$ realizations were made to obtain each curve. The forcing amplitude is low and corresponds to the annual limit cycle, $\beta_1=0.02$. The vertical helper lines mark the frequencies predicted by Eq. (\ref{fr}).} \label{chap6fig3}} 
\end{figure}

\begin{figure}
\centering
{\includegraphics[trim=3.7cm 0.05cm 4.05cm 0.04cm, clip=true, width=\columnwidth]{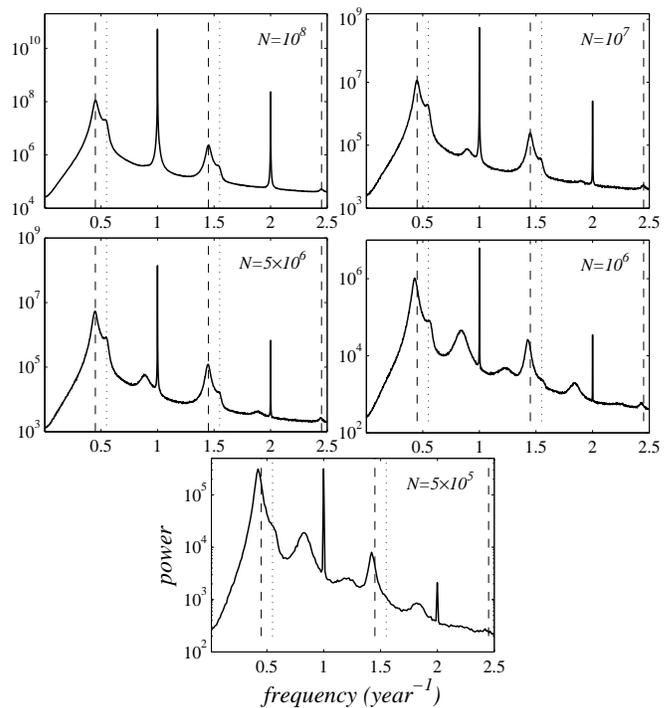}
\vspace{0.1cm}
\caption{Power spectra of the number of infectives calculated from simulations for several system sizes $N$. The simulations of 500 years were run for $N=10^8, 10^7, 5\times10^6, 10^6$ and of 100 years for $N=5\times10^5$. Averages over $10^4$ ($10^3$) realizations were made to obtain the power spectra for $N=10^8$ (for the remaining system sizes). The forcing amplitude is $2.5$ higher than in Fig. \ref{chap6fig3} but the deterministic system is still in the annual limit cycle regime, $\beta_1=0.05$. The vertical helper lines mark the frequencies predicted by Eq. (\ref{fr}).} \label{chap6fig4}} 
\end{figure}

The analytical predictions for stochastic power spectra work well in a large range of $\beta_1$ for annual and biennial limit cycles. However, as the system size is decreased systematic deviations start to appear. We first consider the case of the annual limit cycle induced by a low forcing amplitude. In Fig. \ref{chap6fig3} we compare the power spectra for the number of infective individuals from simulations for several system sizes $N$. In all panels, sharp peaks due to the annual limit cycle can be observed at integer frequencies, as well as several broader stochastic peaks whose frequencies are given by Eq. (\ref{fr}). For all system sizes the first (and largest) stochastic peak is situated at $|\text{Im} \ {\rho_{1,2}}|/(2\pi)$, the second is situated at $1-|\text{Im} \ {\rho_{1,2}}|/(2\pi)$, etc. However, for $N=10^6$ and $N=5\times10^5$ the dominant frequencies of the stochastic peaks are very slightly shifted to the left, so the characteristic period of the stochastic fluctuations gets higher. This effect has already been observed before in the spectra of fluctuations around a stable node, in unforced systems \cite{MSimoes05062008}. We also find that the spectra are fully described by the theory only for very large populations. Beginning with $N=10^7$ a number of much smaller stochastic peaks starts to appear close to the deterministic peaks. Their amplitudes increase as the population size decreases but they stay orders of magnitudes lower than the dominant stochastic and deterministic peaks. These secondary peaks cannot be explained within the theory developed in this study and require considering corrections to the linear Fokker-Planck equation. 
Another effect apparent in small systems is the change in the relative amplitude of the main stochastic and deterministic peaks. For small populations the deterministic limit cycle does not dominate the dynamics of the system any more. The enhancement and broadening of the stochastic peaks show a much noisier and irregular dynamics. For comparison, see the panel for $N=5\times10^5$ in Fig. \ref{chap6fig3} where the main stochastic peak is significantly higher and broader than the main deterministic peak.                

The same picture of changes in the spectra for different population sizes is maintained if the amplitude of forcing is increased but the annual limit cycle is still stable. As an example, see Fig. \ref{chap6fig4} done for $\beta_1=0.05$ (this forcing amplitude is 2.5 larger than the one used in Fig. \ref{chap6fig3}).

\begin{figure}
\centering
{\includegraphics[trim=3.2cm 5.05cm 4.6cm 0.05cm, clip=true, width=\columnwidth]{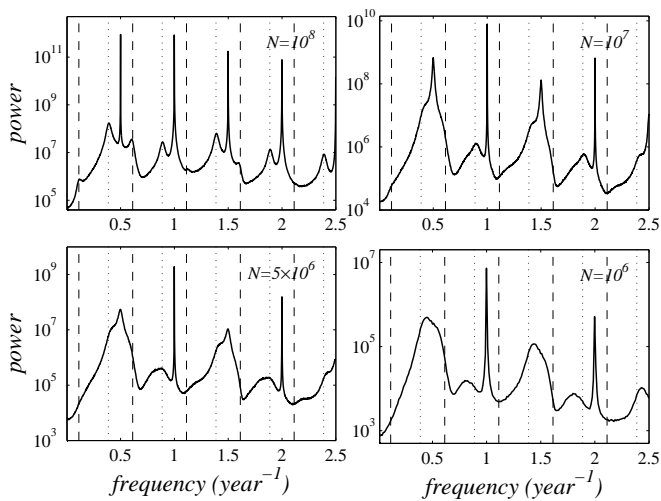}
\vspace{0.1cm}
\caption{Power spectra of the number of infectives calculated from simulations for several system sizes $N$. The simulations of 500 years were run for $N=10^8, 10^7, 5\times10^6$ and of 100 years for $N=10^6$. Averages over $5\times10^3$ ($10^3$) realizations were made to obtain the power spectra for $N=10^8$ (for the remaining system sizes). The forcing amplitude corresponds to the biennial limit cycle, $\beta_1=0.2$. The vertical helper lines mark the frequencies predicted by Eq. (\ref{fr}).} \label{chap6fig5}} 
\end{figure}

If the forcing amplitude is increased even further the period doubling of the limit cycle occurs at $\beta_1'\approx 0.11479$ and everywhere in the interval $0.11479< \beta_1 \leq 0.2$ a stable limit cycle of period 2 is present in the deterministic model. In Fig. \ref{chap6fig5} we compare the spectra from simulations and the analytical spectra for a range of system sizes. Again, the spectra demonstrate narrow deterministic peaks at multiples of $1/2$ due to the limit cycle of period 2 and regular stochastic peaks. For the largest system size, the stochastic peaks are predicted correctly by the theory. The major stochastic peak is now located at $1/2-|\text{Im} \ {\rho_{1,2}}|/(2\pi)$. Note that the peak which used to be dominant in the annual regime was located at $|\text{Im} \ {\rho_{1,2}}|/(2\pi)$. But now this frequency corresponds to the smallest  among all stochastic peaks in the range of frequencies shown in the plot. This observation deserves a longer comment and we will go back to it later. For smaller values of $N$ we observe that the deterministic peak at $1/2$ gets smaller and merges with the two surrounding stochastic peaks. At some point it becomes impossible to distinguish between the stochastic and deterministic peaks which are represented by a single broad peak around $1/2$. The stochastic peaks get significantly enhanced but also the range of frequencies present in the infective time series. 

\begin{figure}
\centering
{\includegraphics[trim=3.9cm 0.cm 4.45cm 0.025cm, clip=true, width=\columnwidth]{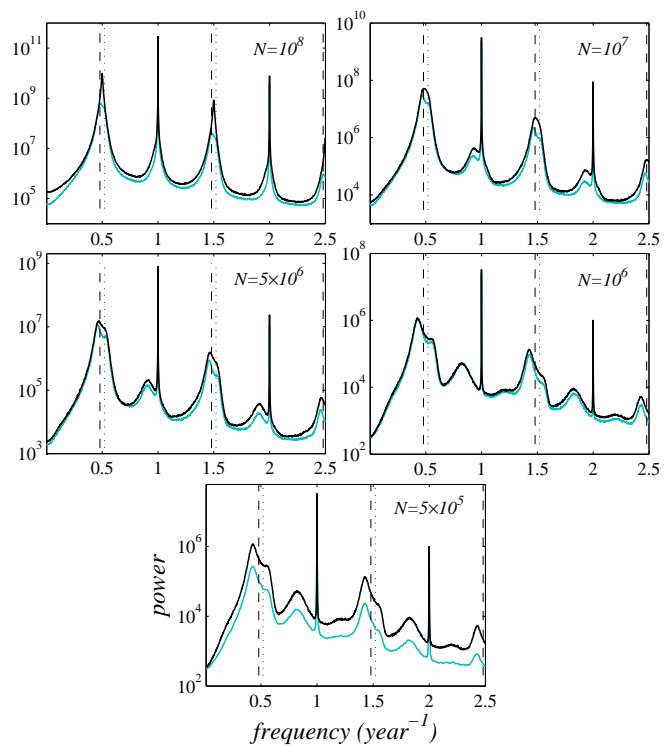}
\vspace{0.1cm}
\caption{(Color online) Power spectra of the number of infectives calculated from simulations for several system sizes $N$. The gray (blue) and black curves were obtained for $\beta_1=0.1$ and $\beta_1=0.12$, respectively. In the deterministic model the period doubling occurs at $\beta_1'\approx 0.11479$. The frequencies of the vertical helper lines correspond to the annual limit cycle. The simulations of 500 years were used for $N=10^8, 10^7, 5\times10^6, 10^6$ and of 100 years for $N=5\times10^5$. Averages over $10^3$ realizations were made to obtain all curves.} 
\label{chap6fig6}} 
\end{figure}

\begin{figure}
\centering
{\includegraphics[trim=3.2cm 7.6cm 4.6cm 0.05cm, clip=true, width=\columnwidth]{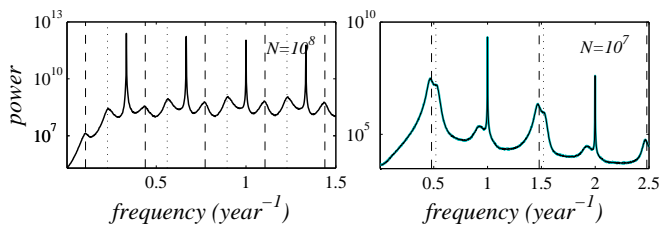}
\vspace{0.1cm}
\caption{(Color online) Power spectra of the number of infectives calculated from simulations for two system sizes $N$ and two sets of initial conditions [for the gray (blue) curve simulations started from random initial conditions and for the black curves the initial conditions were chosen close to the deterministic triennial cycle]. In the left (right) panel the vertical helper lines mark the predicted peak frequencies for the triennial (annual) limit cycle. The amplitude of seasonal forcing corresponds to coexisting stable annual and triennial limit cycles in the deterministic model, $\beta_1=0.1$.} 
\label{chap6fig7}} 
\end{figure}

The next parameter range of interest is close to the point where the period doubling bifurcation occurs for the deterministic system. In
 this regime we do not expect to have an agreement between the theory and simulations because none of the cycles is stable enough, however there are interesting conclusions which can be drawn from the power spectra. Recall that the period doubling in the deterministic model occurs at $\beta_1'\approx 0.11479$ but in the stochastic model the transition from the annual limit cycle to the biennial limit cycle is shifted to a higher value of $\beta_1$. In Fig. \ref{chap6fig6} we compare the spectra for the number of infectives from simulations performed for two close values of $\beta_1$ in the vicinity of the deterministic period doubling point, one before the bifurcation ($\beta_1=0.1$, corresponding to the stable annual limit cycle, gray (blue) curves) and one after ($\beta_1=0.12$, corresponding to the stable biennial limit cycle, black curves). As one can see the transition is blurred and shows up later in the stochastic system. The same behavior of the fluctuations power spectrum around a bifurcation was described in Ref. \cite{wiesenfeld}. There is almost no difference between the simulated spectra. The first deterministic peak signaling the biennial limit cycle has to be present at $1/2$ for the black curves, and this is where the stochastic peaks for the annual limit cycle are located. The two main stochastic peaks of the annual cycle move on to $1/2$ and at some point give rise a deterministic peak of the biennial cycle. This is seen from the plot for the largest system size, however the deterministic peak becomes obvious if the $\beta_1$ is increased a bit further. A similar broadening and shifting of the bifurcation point in simulations has also been observed in the unforced model with a transition from a stable focus to a stable limit cycle through an Andronov-Hopf bifurcation \cite{PhysRevE.79.041922,unforcedmckane}.

The last question one naturally poses is how does the coexistence of attractors influence the power spectrum, through possible switches between basins of attraction induced by stochasticity.
The relevance of this mechanism has been argued for in the literature \cite{Keeling2001317,Lora}, in particular to try to provide an explanation for the diversity of patterns of childhood diseases \cite{Earn01282000, BauchEarn}. In the deterministic model there is a range of $\beta_1$ where stable annual and biennial limit cycles coexist with a stable triennial cycle. In particular, this happens for the parameter values of Fig. \ref{chap6fig6}. We, however, have not identified any deterministic peaks at multiples of $1/3$ associated with a triennial limit cycle or even an indication of such a behavior in the stochastic system. The same was confirmed for other values of $\beta_1$ where the orbits coexist. To clarify this situation we performed an experiment in which one set of simulations started on a triennial limit cycle (these conditions are favorable for the triennial cycle as it will become clear later) and the other set started from a random initial condition. The resulting power spectra are shown in Fig. \ref{chap6fig7} for $\beta_1=0.1$. The black [respectively, gray (blue)] curves are the power spectra of the infective time series beginning from favorable (respectively, random) conditions. It is only for system size $N=10^8$ that we were able to observe a triennial limit cycle with the fluctuations around it described by the van Kampen expansion (see the left panel in Fig. \ref{chap6fig7}, the helper lines mark the frequencies for the triennial limit cycle). For all other system sizes the power spectra are identical to those already shown in Fig. \ref{chap6fig6}.

\begin{figure}
\centering
{\includegraphics[trim=2.7cm 0.cm 5.cm 0.cm, clip=true, width=\columnwidth]{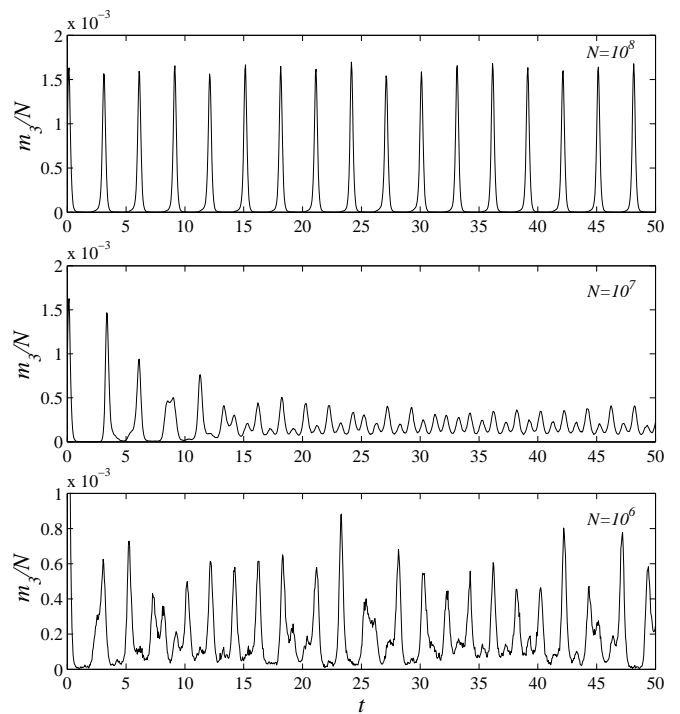}
\vspace{0.1cm}
\caption{The infective density recorded from a typical realization of the stochastic model starting on the deterministic triennial limit cycle. The seasonal forcing amplitude is $\beta_1=0.1$.} 
\label{chap6fig8}} 
\end{figure}

\begin{figure}
\centering
{\includegraphics[trim=2.7cm 0.1cm 5.cm 0.cm, clip=true, width=\columnwidth]{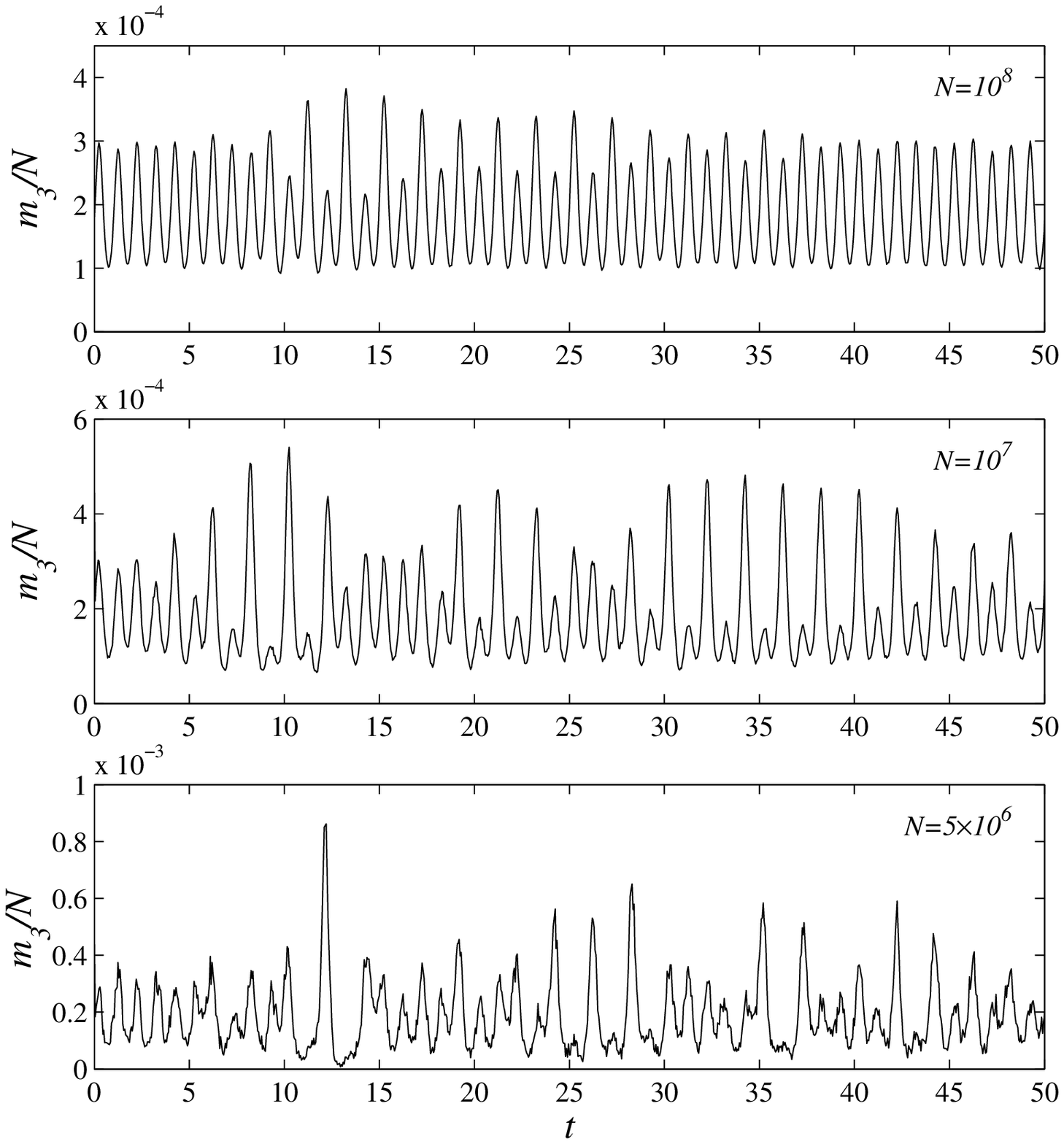}
\vspace{0.1cm}
\caption{The infective density recorded from a typical realization of the stochastic model starting on the deterministic annual limit cycle. The seasonal forcing amplitude is $\beta_1=0.1$.} 
\label{chap6fig9}} 
\end{figure} 

\begin{table*}
\centering
{\begin{tabular}{|c|p{4.5cm}|p{4.5cm}|p{4.5cm}|}
\hline $\beta_1$ & \hspace{1.65cm}$N=10^8$ & \hspace{1.4cm}$N=10^7$ & \hspace{0.7cm}$5\times10^5\leq N\leq 10^6$\\
\hline \multirow{4}{*}{$0.05$} & well defined dominant annual deterministic peak & well defined dominant annual deterministic peak and two subdominant close to biennial broad stochastic peaks& well defined annual deterministic peak and close to biennial broad stochastic peak \\
\hline 
 \multirow{4}{*}{$0.2$} & well defined dominant annual and biennial deterministic peaks & well defined dominant annual deterministic peak and subdominant broad biennial stochastic peak & well defined dominant annual deterministic peak and subdominant close to biennial much broader stochastic peak\\ \hline
\end{tabular}
\caption{Qualitative features of the power spectra for different values of $\beta_1$ and $N$.}
\label{table3}}
\end{table*}

In this case, a better insight into what happens in the simulations is given by inspecting the time series. In Fig. \ref{chap6fig8} we show the density of infectives in a typical single realization of the stochastic model starting from the triennial limit cycle (for the same value of $\beta_1$ as in Fig. \ref{chap6fig7}). For system size $N=10^8$ the density exhibits regular triennial oscillations. Their characteristic features are a high amplitude (at least twice as large as that of a typical annual or biennial cycle) and very low number of infectives between recurrent epidemics. For $N=10^8$ the system tends to stay on the triennial limit cycle because the fluctuations are not large enough to drive it to the annual limit cycle. But for smaller $N$ the system either goes extinct or drops very quickly to the annual limit cycle (or a biennial one if $\beta_1 > 0.11479$) and stays on it. For the population of one million of individuals not even one oscillation over the triennial cycle is followed by the stochastic system because the relative fluctuations are large at the first drop of infectives and drive the system to the basin of attraction of the annual cycle. 
In Fig. \ref{chap6fig9} we show some of the time series for the same parameter values as in the previous figure but for initial conditions starting on the annual limit cycle. The density exhibits alternating regions of annual and biennial oscillations for large population sizes with an ever increasing role of the stochastic fluctuations as $N$ decreases. For small sizes the time series look quite irregular, occasionally exhibiting interepidemic intervals longer than 2 years. This case corresponds to the power spectra where the stochastic peaks are broad and high. 
It is remarkable that we have never observed the backward switching from an annual (or biennial) limit cycle to a triennial one.

The qualitative features of the power spectra for two different values of the forcing amplitude, $\beta_1$, and different population sizes, $N$, are summarized in Table \ref{table3}. As discussed above, well defined high amplitude peaks show up in the time series as regular cycles, and broad stochastic peaks as lower amplitude noisy oscillations.

\section{V. Discussion and conclusions}

In conclusion, in this study we have considered the deterministic and the stochastic SEIR models with sinusoidally varying contact rate. Depending on the forcing amplitude, the deterministic model exhibits a range of stable attractors, the most visible of which are limit cycles of periods 1, 2 and 3. Using the van Kampen's expansion of the master equation of the corresponding stochastic model we have calculated autocorrelation functions and power spectra of the stochastic fluctuations around these attractors. We have compared the analytical results with those obtained from direct simulations of the stochastic model. We have found that in a large range of the forcing amplitude there is an excellent agreement between the theory and simulations. The prerequisites for this are large system sizes (typically higher than $10^6$) and the stability of attractors, namely the more stable the cycle and the higher the system size are the better the agreement between the simulated and analytical power spectra is. This is exactly what we would expect for the quality of the approximation given by the truncation at first order of the  full van Kampen expansion. 

The power spectra of the infective time series demonstrate peaks of two types. The narrow peaks are due to a limit cycle of a given period and the broader peaks are due to the resonant amplification of stochastic fluctuations around the limit cycle. It has been argued that the presence and position of deterministic and stochastic peaks in the power spectra obtained from real data records of childhood diseases can be predicted by the deterministic theory alone, and that, moreover, the frequency of the stochastic peak is defined by the frequency of the transients near a stable limit cycle \cite{BauchEarn,Bausch,RohaniKeelingGrenfell}. We have identified the main frequencies of the stochastic peaks in the annual and biennial regimes and shown that these do not necessarily equal the frequency of the damped oscillations of deterministic perturbations around a cycle. Thus, neither the full structure of the power spectrum of the infective time series nor the most prominent frequency of the stochastic peak can be fully predicted by the deterministic model.

Another conclusion concerns the role of the coexistence of attractors in the seasonally forced stochastic and deterministic SEIR models. The coexistent stable limit cycles present in the deterministic epidemic models have been conjectured to be the reason why irregular dynamics is observed in the corresponding stochastic system. Namely, it has been a systematic assumption of several papers on childhood infectious diseases modeling \cite{Earn01282000,schwartz1985,BauchEarn} based exclusively on the deterministic analysis that for small populations the stochastic system should constantly switch between cycles of different periods due to demographic or environmental noise, thus exhibiting irregular dynamics. The results of the present study contradict this view. For values of $\beta_1$ close to the period doubling bifurcation value, demographic as well as environmental noise can promote switching between period 1 and period 2 cycles. However, for the stable limit cycles of period 1 or 2 coexisting with a stable cycle of period 3 such switching is absent in the stochastic system. Notwithstanding the basin of attraction of the triennial limit cycle being roughly $25\%$ of the total initial conditions \cite{schwartz1985} for the deterministic system, it is extremely unstable in the simulations. Even for large system sizes the system is brought to a cycle of lower period after a very short time, and backwards switching is not observed. This can be understood from the shape of the periodic solutions of the triennial cycle of the deterministic model, together with the fact that most initial conditions with very low number of infectives are attracted to the lower period cycles. 

We have become aware of a recent paper where the fluctuation power spectrum of a seasonally forced stochastic SIR model is computed through the same method \cite{epidforcingmckane}. In this paper,
a transition representing injection of infectives is included in addition to the processes of infection, recovery, and birth-death. It is shown that, even for very low infective immigration rates, the inclusion of this term has a drastic impact on the bifurcation diagram of the deterministic model, leaving stable annual and biennial limit cycles as the only possible attractors \cite{epidforcingmckane}. This shows that the competing higher period attractors  of the forced system are very fragile indeed. They are not robust, in the deterministic setting, with regard to small changes in the model. On the other hand, in the stochastic setting, their effective basin of attraction is negligibly small, except for unrealistically large population sizes. 
The results of this study and those of Ref. \cite{epidforcingmckane} concur to support the view that the main ingredient to understand the observed incidence patterns of childhood infectious diseases is stochastic amplification, rather than noise induced switching between competing attractors of the deterministic system.

\textbf{\begin{center}
Acknowledgments
\end{center}}
Financial support from the Foundation of the University of Lisbon (FUL) 
and the Portuguese Foundation for Science and 
Technology (FCT) under Contract No. POCTI/ISFL/2/261 is gratefully 
acknowledged. The first author (G.R.) was also supported by FCT under Grant No. SFRH/BD/32164/2006 and by Calouste Gulbenkian Foundation under its Program "Stimulus for  Research".

\section{Appendix}
As described in the Section III of the main text, the master equation for the seasonally forced stochastic SEIR model is constructed by considering all possible transitions increasing or decreasing the probability of finding a system in state with $m_1$ susceptible, $m_2$ exposed and $m_3$ infective individuals. This equation reads as follows: 
 
\begin{widetext}
\begin{eqnarray}
\label{masterseasonal}
\dfrac{d\mathcal{P}(m_1,m_2,m_3,t)}{dt}&=&\mathcal{T}^{m_1,m_2,m_3}_{m_1+1,m_2-1,m_3}\mathcal{P}(m_1+1,m_2-1,m_3,t)+\mathcal{T}^{m_1,m_2,m_3}_{m_1,m_2,m_3+1}\mathcal{P}(m_1,m_2,m_3+1,t) \ \ \  \nonumber\\
&+&\mathcal{T}^{m_1,m_2,m_3}_{m_1,m_2+1,m_3-1}\mathcal{P}(m_1,m_2+1,m_3-1,t)+\mathcal{T}^{m_1,m_2,m_3}_{m_1-1,m_2,m_3}\mathcal{P}(m_1-1,m_2,m_3,t)\nonumber\\
&+&\mathcal{T}^{m_1,m_2,m_3}_{m_1+1,m_2,m_3}\mathcal{P}(m_1+1,m_2,m_3,t)+\mathcal{T}^{m_1,m_2,m_3}_{m_1,m_2+1,m_3}\mathcal{P}(m_1,m_2+1,m_3,t)\nonumber\\&+&\mathcal{T'}^{m_1,m_2,m_3}_{m_1,m_2,m_3+1}\mathcal{P}(m_1,m_2,m_3+1,t) -\left[\vphantom{\mathcal{T'}^{m_1,m_2,m_3-1}_{m_1,m_2,m_3}}\mathcal{T}^{m_1-1,m_2+1,m_3}_{m_1,m_2,m_3}+\mathcal{T}^{m_1,m_2-1,m_3+1}_{m_1,m_2,m_3}\right.\nonumber\\
&+&\mathcal{T}^{m_1,m_2,m_3-1}_{m_1,m_2,m_3}+\mathcal{T}^{m_1+1,m_2,m_3}_{m_1,m_2,m_3}+\mathcal{T}^{m_1-1,m_2,m_3}_{m_1,m_2,m_3}+\mathcal{T}^{m_1,m_2-1,m_3}_{m_1,m_2,m_3}\nonumber\\
&+&\left.\mathcal{T'}^{m_1,m_2,m_3-1}_{m_1,m_2,m_3}\right]\mathcal{P}(m_1,m_2,m_3,t).
\end{eqnarray}
\end{widetext}

The above equation and its subsequent analysis can be simplified with the use of the step operators defined in the main text by Eqs. (\ref{epsilon1seas})-(\ref{epsilon3seas}). The substitution of Eqs. (\ref{epsilon1seas})-(\ref{epsilon3seas}) into Eq. (\ref{masterseasonal}) gives 
  
\begin{widetext}
\begin{eqnarray}
\label{seasmastlast}
\dfrac{d\mathcal{P}(m_1,m_2,m_3,t)}{dt}&=&\Big[\left(\epsilon_2\epsilon_3^{-1}-1\right)\mathcal{T}^{m_1,m_2-1,m_3+1}_{m_1,m_2,m_3}\Big.+\left(\epsilon_1-1\right)\mathcal{T}^{m_1-1,m_2,m_3}_{m_1,m_2,m_3}+\left(\epsilon_3-1\right)\mathcal{T}^{m_1,m_2,m_3-1}_{m_1,m_2,m_3}\nonumber \ \ \ \\
&+&\left(\epsilon_1\epsilon_2^{-1}-1\right)\mathcal{T}^{m_1-1,m_2+1,m_3}_{m_1,m_2,m_3}+\left(\epsilon_1^{-1}-1\right)\mathcal{T}^{m_1+1,m_2,m_3}_{m_1,m_2,m_3}+\left(\epsilon_2-1\right)\mathcal{T}^{m_1,m_2-1,m_3}_{m_1,m_2,m_3}\nonumber\\
&+&\Big.\left(\epsilon_3-1\right)\mathcal{T'}^{m_1,m_2,m_3-1}_{m_1,m_2,m_3}\Big]\mathcal{P}(m_1,m_2,m_3,t).
\end{eqnarray}
\end{widetext}


\begin{thebibliography}{99}
\bibitem{comment}
For the purposes of fitting to measles incidence time series, the term-time forced SEIR model is sometimes preferred. The sinusoidally forced SEIR model, however, reproduces the behavior of the term-time forced SEIR model at a lower level of forcing (see the comment (10) in Ref. \cite{Earn01282000}). Here we use Eq. (\ref{seasonalrate}) as it will somewhat simplify the analysis of stochastic effects in the SEIR model with a periodically varying contact rate still allowing us to understand the main mechanisms at play.
\end{thebibliography}
\end{document}